# *Putting Privacy into Perspective - Comparing Technical, Legal, and Users' View of Data Sensitivity*

Eva-Maria Schomakers[1], Chantal Lidynia[1], Dirk Müllmann[2], Roman Matzutt[3], Klaus Wehrle[3], Indra Spiecker gen. Döhmann[2], Martina Ziefle[1]

[1] RWTH Aachen University, Germany, {schomakers|lidynia|ziefle}@comm.rwth-aachen.de

[2] Goethe University, Frankfurt am Main, Germany, {muellmann|spiecker}@jur.uni-frankfurt.de

[3] RWTH Aachen University, Germany, {matzutt|wehrle}@comsys.rwth-aachen.de

**Abstract:** Web 2.0, social media, cloud computing, and IoT easily connect people around the globe, overcoming time and space barriers, and offering manifold benefits. However, the technological advances and increased user participation generate novel challenges for protecting users' privacy. From the user perspective, data disclosure depends, in part, on the perceived sensitivity of that data, and thus on a risk assessment of data disclosure. But in light of the new technological opportunities to process and combine data, it is questionable whether users are able to adequately evaluate the risks of data disclosures. As mediating authority, data protection laws try to protect user data, granting enhanced protection to "special categories" of data. In this publication, the legal, technological, and user perspectives on data sensitivity are presented and compared. From a technological perspective, all data can be referred to as "potentially sensitive." The legal and user perspective on data sensitivity deviate as some data types are granted special protection by the law but are not perceived as very sensitive by the users, and vice versa. Merging the three perspectives, the implications for informational self-determination are discussed.

## I. Introduction

The early 2000's shift of online services towards the Web 2.0 paradigm[1] constituted a revolution of online services: user participation became an elementary ingredient of modern online services to enable, e.g., more direct forms of communication or provide personalized and thus improved services to their users. More and more technologies that offer convenience, entertainment, and support for users in almost all areas of life evoke that a large proportion of everyday life of typical internet users takes place using online services. For that and thereby, users generate and disclose large amounts of data.

But users are not always content with that: many users are concerned about their online privacy and data collection[2]. Nevertheless, the so-called *privacy paradox* describes the phenomenon that users still disclose much personal data despite their stated concerns[3]. They often feel forced to disclose more data than they actually would like to[4] and are at the same time unsure about their rights and privacy protection possibilities[5]. For example, 58% of Europeans incorrectly believe that chat messages are confidential by law.[6]

Users concerns and paradoxical sharing behavior encounters developing technological possibilities. Data collection and processing has evolved over time so that increasingly more information can be combined, deanonymized, and used to profile individuals. For example, the recent Cambridge Analytica scandal showed how this user data and user profiles can even be used to influence democratic elections. And the data used by Cambridge Analytica was voluntarily disclosed by the users. Here, the question arises again, why users disclose data when they are concurrently worried about its use? Many aspects have been considered to explain privacy paradoxical behaviors, e.g., missing awareness or knowledge, convenience and benefits for data disclosure that override perceived risks, or heuristics and cognitive limitations that influence decision behaviors.[7] One central

---

[1] Tim O'Reilly, 'What Is Web 2.0: Design Patterns and Business Models for the Next Generation of Software' [2007] International Journal of Digital Economics 17

[2] European Commission, *Special Eurobarometer 447 – Online Platforms* (2016)

[3] Nina Gerber, Paul Gerber, Melanie Volkamer, 'Explaining the Privacy Paradox: A Systematic Review of Literature Investigating Privacy Attitude and Behavior' (2018) 77 Computers and Security 226

[4] Sabine Trepte, Philipp K Masur, 'Privacy Attitudes, Perceptions, and Behaviors of the German Population: Research Report' in Michael Friedewald and others (eds), *Forum Privatheit und selbstbestimmtes Leben in der digitalen Welt* (2017)

[5] Sabine Trepte, Philipp K Masur, 'Privacy Attitudes, Perceptions, and Behaviors of the German Population: Research Report' in Michael Friedewald and others (eds), *Forum Privatheit und selbstbestimmtes Leben in der digitalen Welt.* (2017); European Comission, *Flash Eurobarometer 443: Report e-Privacy* (2016)

[6] European Comission, *Flash Eurobarometer 443: Report e-Privacy* (2016)

[7] Alessandro Acquisti, Laura Brandimarte, George Loewenstein, 'Privacy and Human Behaviour in the Age of Information' (2015) 347 Science 509; Nina Gerber, Paul Gerber, Melanie Volkamer, 'Explaining the Privacy Paradox: A Systematic Review of Literature Investigating Privacy Attitude and Behavior' (2018) 77 Computers

aspect is that users evaluate the risks of data disclosure and that this evaluation is biased. How risky data disclosure is perceived, and correspondingly, how willingly it is disclosed, is strongly influenced by the type of data at stake -- and its perceived sensitivity.[8] But in light of the new technological possibilities that introduce increasing potential to data use, this sensitivity evaluation may not be up-to-date.

Undoubtedly, there are, nonetheless, many occasions in which the disclosure and processing of personal data is indicated and necessary. E.g., if an e-commerce company is to sell articles, they have to obtain an address where to ship to and also receive payment, therefore having to gain access to bank data and charge the wares worth. Also, many digital services are offered without charge in exchange for the user data. In these situations, a reasonable and responsible handling of user data is necessary to most online services and transactions. As a mediating authority between the commercial demand and the privacy concerns, there are laws that govern the use of user data by companies. In Europe, data protection is for the most part regulated by the new European General Data Protection Regulation (GDPR), which distinguishes between different categories of information that are granted different levels of protection. The GDPR concerns only personal data that is defined as information relating to an identified or identifiable natural person. It also knows special categories of personal data embracing particularly sensitive information about natural persons. But with the ever-improving potential for data analysis the question arises whether the regulation (legal perspective) captures what can potentially become sensitive (technological perspective) and also what users perceive to be sensitive data (user perspective).

The aim of the GDPR is to strengthen the right to informational self-determination, thus guaranteeing a save environment in which users can decide freely what services to use and what data to share. In this paper, we will examine sensitivity of information from the three perspectives, legal, technological, and user, and compare these perspectives. We will begin with providing an overview of how data collection and processing by online services has evolved over time and how new technologies redefine what must be considered potentially sensitive data. For the legal perspective, we discuss how the GDPR distinguishes between different categories of data and what protection it grants these. To contribute a user perspective on information privacy, n = 601 German internet users evaluated the perceived sensitivity of 40 different data types in an online questionnaire. This data forms the basis of a comparison between the three perspectives and a discussion of the findings.

---

and Security 226; Lemi Baruh, Ekin Secinti, Zeynep Cemalcilar, 'Online Privacy Concerns and Privacy Management: A Meta-Analytical Review' (2017) 67 Journal of Communication 26

[8] David L Mothersbaugh and others, 'Disclosure Antecedents in an Online Service Context: The Role of Sensitivity of Information' (2012) 15 Journal of Service Research 76

## II. Information Sensitivity from a Technological Perspective

We begin our discussion by reenacting how user interaction on the Internet evolved over time, how this development yielded new opportunities as well as severe privacy risks, and how technology aims to overcome these challenges. This discussion subsequently serves as a foundation to assess the users' and legal perspective on (perceived) data sensitivity. We sketch how paradigm shifts promoted higher degrees of user interaction and sharing of data and how recent technologies support this trend.

We then complement these observations with a discussion of privacy challenges stemming from the technological development, where we conclude that previously uncritical data can potentially **become** highly sensitive given technological advances and that single technical countermeasures cannot mitigate all arising risks.

### 1. Evolving User Interaction on the Internet

The early 2000's shift of online services towards the Web 2.0 paradigm[9] constituted a revolution of online services: user participation became an elementary ingredient of modern online services to enable, e.g., more direct forms of communication or provide personalized and thus improved services to their users. Prominent examples of classical Web 2.0 services comprise knowledge databases such as Wikipedia, media sharing platforms such as YouTube, Flickr, or SoundCloud, and personal blogs as a user-friendly evolution of personal homepages.[10] However, also the utilization of user statistics, e.g., their search queries or surfing behavior, became more important once user participation was shifted to the center of attention as they allow optimizing the user experience.[11]

The level of user interaction culminated in the rise of **social networks** such as Twitter and Facebook, which brings users and companies closer together in a highly interactive manner: users can create links to their friends and relatives and publicly or privately share memorable moments or opinions in text, picture, video, or by sharing or commenting other users' posts. In consequence, social networks still remain popular today, as is expressed by Facebook's 2.3 billion and Twitter's 335 million monthly active users, respectively.[12]

This rise of such global platforms was enabled by shifting to the **cloud computing paradigm**[13], which enabled service providers to rapidly tailor computing resources to their individual needs with a drastically lowered need for technical expertise. However, users did not only implicitly benefit from the cloud via its increased scalability and lowered entry barrier for service providers. The cloud's ubiquity also enabled users to outsource their data to the cloud in order to simplify sharing data or maintaining online backups, as showcased by the popular services provided by, e.g., Dropbox, Google Drive, or Amazon Drive.

---

[9] Tim O'Reilly, 'What Is Web 2.0: Design Patterns and Business Models for the Next Generation of Software' [2007] International Journal of Digital Economics 17
[10] Ibid.
[11] Ibid.
[12] Sara Salinas, 'Peak Social? The Major Social Platforms Are Showing a Significant Slowdown in Users' < https://www.cnbc.com/2018/08/08/social-media-active-users-around-the-world.html> accessed June 2019
[13] Michael Armbrust and others, 'A View of Cloud Computing' (2010) 53 Commun. ACM 50

Ultimately, the advent of **smartphones** and the **Internet of Things (IoT)**, e.g., in the form of fitness trackers or smart home devices, would have been unthinkable without the cloud. On the one hand, cloud storage enabled online services to synchronize different devices of users very easily. Hence, users can now access their relevant data on both their smartphones and their notebooks or desktop PCs with minimal management overhead. On the other hand, the cloud's scalable processing power enables service providers to remotely process the data sensed by their users' IoT devices, which are typically resource-constrained and thus unable to perform costly computations.[14] However, the inherent centrality of the cloud is also source of user reservations as we detail in the next section.

As a consequence, systems based on **distributed ledgers**, most notably blockchains, recently gained traction to break up this level of centralization: while initial blockchain-based systems such as Bitcoin[15] or Ethereum[16] focused on achieving trustless distributed financial services, distributed ledgers are now being explored to be used for, e.g., tamper-proof file storage[17] or access control management of user data,[18] i.e., services concerning data that is currently being hosted on the cloud. Distributed ledgers are experiencing this popularity because their immutability establishes technical accountability among otherwise mutually distrusting parties.

In conclusion, new technologies always simplified the deployment of online services centered around user participation or even enabled novel services over the Internet. However, those opportunities do not come without additional challenges, which we discuss in the subsequent section.

## *2. Increasing Data Sensitivity*

In the preceding section, we sketched how the Web 2.0 encouraged all users to participate in the content creation, yielding novel Internet services such as social media. Technological advances such

---

[14] Martin Henze and others, 'Maintaining User Control While Storing and Processing Sensor Data in the Cloud' (2013) 5 *International Journal of Grid and High Performance Computing (IJGHPC)*

[15] Satoshi Nakamoto, 'Bitcoin: A Peer-to-Peer Electronic Cash System' (2008) <https://bitcoin.org/bitcoin.pdf> accessed June 2019

[16] Gavin Wood, 'Ethereum: A Secure Decentralised Generalised Transaction Ledger' (2016) <http://gavwood.com/Paper.pdf> accessed June 2019

[17] Henning Kopp and others, 'Design of a Privacy-Preserving Decentralized File Storage with Financial Incentives', 2017 *IEEE European Symposium on Security and Privacy Workshops (EuroS&PW)* (2017)

[18] Guy Zyskind, Oz Nathan, Alex Pentland, 'Decentralizing Privacy: Using Blockchain to Protect Personal Data', *2015 IEEE Security and Privacy Workshops* (2015)

as cloud computing, smartphones, wearables and smart home raised the shareability and availability of user data on the Internet to new dimensions. Furthermore, recent advances such as distributed ledgers are currently being explored as a potential next step in this vein. In this section, we argue that all these technological advances come with novel challenges for protecting users' privacy. Most notably, they can cause new types of user data to **become** sensitive either due to new kinds of data being exposed online or due to advanced analysis capabilities deriving new insights from the available data. Hence, it is thus more appropriate to refer to all user data on the Internet as **potentially sensitive data** as we discuss in the following:

**Web2.0.** The shift towards centering around user participation within the Web 2.0 paradigm inherently led to the collection of more user data. This is problematic with respect to insights gained from **service personalization**, **user tracking**, and **data breaches.** Online services are routinely personalized to increase the user experience, e.g., provide individually better search results. While personalized services can benefit the user, the collected data is potentially highly sensitive. For instance, in 2006 AOL infamously released users' search queries from their search engine for research purposes. However, only the user identifiers were anonymized, and the search queries were left untouched. Journalists were thus able to easily deanonymize users solely based on their query history.[19] It is further possible via more advanced analysis methods to disclose sensitive user data even from properly anonymized data sets. For example, in 2006 Netflix released a data set of anonymous movie ratings, which could be disclosed by correlating the anonymized data with movie ratings from IMDb even for small overlaps of rated movies in both data sets.[20] The authors claim that this, for instance, can potentially reveal additional latent attributes of the users such as their political views.[21] Web tracking maximizes this form of data collection by monitoring users' browsing behavior **across** services.[22] Knowing the set of visited websites or even, for instance, individual online articles, potentially discloses a much more fine-grained view on users compared to their behavior interacting with a single service. Furthermore, web tracking is oftentimes technically opaque to the user, which is why today legislation begins to counter such practices, e.g., the European GDPR now mandates service operators to inform users about their utilization of tracking cookies. Yet, even privacy-aware users actively protecting their privacy by deleting cookies or using private browsing have been shown

---

[19] Michael Barbaro, Tom Zeller Jr., 'A Face Is Exposed for AOL Searcher No. 4417749' <https://www.nytimes.com/2006/08/09/technology/09aol.html> accessed June 2019

[20] Arvind Narayanan, Vitaly Shmatikov, 'Robust De-Anonymization of Large Sparse Datasets', *2008 IEEE Symposium on Security and Privacy (SP 2008)* (2008)

[21] Ibid.

[22] Jonathan R Mayer, John C Mitchell, 'Third-Party Web Tracking: Policy and Technology', *Proceedings of the 2012 IEEE Symposium on Security and Privacy (IEEE Computer Society 2012)*

to be susceptible to web tracking due to their distinct behavior.[23] While service personalization and web tracking disclose sensitive user data explicitly or implicitly to the respective service operators, a major threat also lies in the potential for data breaches. Such breaches typically release login credentials and other meta data for users' accounts registered with the attacked platform. At the time of writing, the website ``Have I Been Pwned?'' which enables users to check whether or not their credentials were affected by a certain data breach, lists 324 of such breaches, affecting 5.5 billion records in total.[24]

**Social Media.** Especially the rise of social media revolutionized users' online behavior as they allowed them to rapidly share personal moments and thoughts with both their friends and a general audience. However, this new-found freedom also came along with unprecedented privacy issues due to sensitive data disclosure. On the one hand, users can directly share clearly sensitive data with the public. For instance, some users posted private credit card information on Twitter, which were collected by one Twitter account in the past.[25] On the other hand, users can release sensitive information via meta data such as GPS locations stored in uploaded pictures.[26] Social networks can aggravate the consequences of such disclosures in cases when users disclose information of another user or when a user is unaware of their (effective) privacy settings.[27] These incidents showcase the need for users to be educated about potential threats arising from sharing sensitive data on the Internet and that users must carefully gauge which data to share and which not to share. Hence, the manifold opportunities of expressing themselves via social media mandates that users educate themselves about technical implications and potential consequences of using corresponding services, ultimately gauge carefully which data to share and which not to share. Furthermore, the data users share online can be combined and subsequently collectively be exploited as shown by the Cambridge Analytica scandal.[28] It was reported that Cambridge Analytica obtained profile data of 30 million Facebook users without their consent from which they were able to create psychographic profiles, e.g., by incorporating other data sources and matching that data to the Facebook users.[29] This enabled the corporation to derive sensitive attributes such as political views, life satisfaction, or religious beliefs from the collected data and subsequently use it to steer affected users via targeted advertisements.[30] Hence, users can be profiled based on their shared data and the increased potential stemming from new analysis methods to exploit such data can cause such data to effectively become sensitive.

---

[23] Ting-Fang Yen and others, 'Host Fingerprinting and Tracking on the Web: Privacy and Security Implications', *The 19th Annual Network and Distributed System Security Symposium (NDSS) 2012 (19th Annu. Netw. Distrib. Syst. Secur. Symp. 2012, Internet Society 2012)*

[24] Troy Hunt, 'Have I Been Pnwed?' (2013) <https://haveibeenpwned.com> accessed June 2019

[25] Twitter user @NeedADebitCard, 'Debit Card | Twitter' (2012) <https://twitter.com/needadebitcard> accessed June 2019

[26] Matthew Smith and others, 'Big Data Privacy Issues in Public Social Media', *2012 6th IEEE International Conference on Digital Ecosystems and Technologies (DEST)* (2012)

[27] Balachander Krishnamurthy, Craig E Wills, 'On the Leakage of Personally Identifiable Information via Online Social Networks', *Proceedings of the 2Nd ACM Workshop on Online Social Networks* (ACM 2009)

[28] Matthew Rosenberg, Nicholas Confessore, Carole Cadwalladr, 'How Trump Consultants Exploited the Facebook Data of Millions' (2018) <https://www.nytimes.com/2018/03/17/us/politics/cambridge-analytica-trump-campaign.html> accessed June 2019

[29] ibid

**Cloud Computing.** Users outsourcing data to the cloud can cause a perceived loss of control over their data. On the one hand, due to the cloud's multitenancy, single cloud providers could gain access to user data of all customers.[31] This is especially highlighted by the high prevalence of cloud services being used to fuel, e.g., smartphone applications[32] or email services,[33] which may be opaque to the user. On the other hand, computing clouds can be distributed over multiple data centers. In this case, users lose control over **where** their data is being stored, which can violate both private and even legal requirements.[34] Hence, the increased technical complexity of data management must be addressed in order to enable users to make educated decisions about outsourcing potentially sensitive data to online services.

**Smartphones and IoT.** The ubiquity of sensing devices such as smartphones or IoT devices, including wearables such as fitness trackers, create new challenges for user privacy.[35] To a large extent, this stems from the fact that third parties can potentially extract very fine-grained information from a user's sensor data using appropriate analysis technologies. While, for instance, traditional location-based services (e.g., on a user's smartphone) only report single locations, other services process whole trajectories,[36] e.g., GPS tracks generated by a user's fitness tracker, which often allows for a much more fine-grained tracking of users. Furthermore, the often-insufficient security of IoT devices for smart homes can potentially leak sensitive information directly from the user's house to the Internet.[37] This potential threat is further exemplified by recent advances in **deep learning.[38]** For example, China is now technically able to track citizens in public spaces via a large network of governmental IoT devices, which is used to denounce, e.g., speeders publicly.[39] In conclusion, in addition to potentially having her data being exposed to third parties without being in control, the user must also consider the potentials of involved analyses of her sensed data to reveal additional sensitive information about her. This especially holds for the envisioned paradigm of **data markets**,[40] where users could actively sell sensor data to interested analysts.

---

[30] Ibid.

[31] Martin Henze and others, 'A Trust Point-Based Security Architecture for Sensor Data in the Cloud' in Helmut Krcmar, Ralf Reussner and Bernhard Rumpe (eds), *Trusted Cloud Computing* (Springer 2014)

[32] Martin Henze and others, 'CloudAnalyzer: Uncovering the Cloud Usage of Mobile Apps', *Proceedings of the 14th EAI International Conference on Mobile and Ubiquitous Systems: Computing, Networking and Services (MobiQuitous)* (2017)

[33] Martin Henze, Mary Peyton Sanford, Oliver Hohlfeld, 'Veiled in Clouds? Assessing the Prevalence of Cloud Computing in the Email Landscape', *2017 Network Traffic Measurement and Analysis Conference (TMA)* (2017)

[34] Martin Henze, Rene Hummen, Klaus Wehrle , 'The Cloud Needs Cross-Layer Data Handling Annotations', *2013 IEEE Security and Privacy Workshops* (2013)

[35] Jan Henrik Ziegeldorf, Oscar Garcia Morchon, Klaus Wehrle , 'Privacy in the Internet of Things: Threats and Challenges' (2013) 7 Security and Communication Networks 2728

[36] Jan Henrik Ziegeldorf and others, 'TraceMixer: Privacy-Preserving Crowd-Sensing sans Trusted Third Party', *2017 13th Annual Conference on Wireless On-demand Network Systems and Services (WONS)* (2017)

[37] Martin Serror and others, 'Towards In-Network Security for Smart Homes', *Proceedings of the 13th International Conference on Availability, Reliability and Security* (ACM 2018)

[38] Yann LeCun, Yoshua Bengio, Geoffrey Hinton, 'Deep Learning' (2015) 521 Nature 436

**Distributed Ledgers.** A potential shift towards distributed ledger-based online service also increases the potential sensitivity of data due to the ledgers' often public nature and their immutability by design. While privacy-preserving platforms based on distributed ledgers are being explored that aim to mitigate public data disclosure[41] sensitive data on the blockchain can have especially devastating consequences. On the one hand, the initial promise of blockchains to provide financial privacy has been falsified by multiple works concerning user anonymity in Bitcoin.[42] On the other hand, it has been shown that arbitrary data can be stored directly on blockchains.[43] This can have serious privacy issues as users tend to store memorable private moments on the blockchain, culminating in reported incidents of doxing on the Bitcoin blockchain, i.e., the malicious and in this case irrevocable disclosure of sensitive data of another user.[44] These initial observations with respect to the emerging distributed ledger technology indicate that users once again will be facing increasing complexity in the technologies fueling their online services in the future.

In conclusion, emerging new technologies can cause intuitively non-sensitive data to become sensitive due to the potential unwanted impact seizing this data can have. Despite whole research areas being dedicated to protect sensitive data from being disclosed to unauthorized parties, there is no general solution to technical data protection. This is exemplified by the fact that today's privacy-enhancing technologies such as secure multiparty computation are typically heavily tailored to only a class of problems.[45] We expect this effect to be further accelerated in the future as widespread technologies become more complex and diverse, and thus harder for users to keep track of in order to gauge potential privacy threats, as we discuss in the subsequent section.

---

[39] Paul Mozur, 'Inside China's Dystopian Dreams: A.I., Shame and Lots of Cameras' <https://www.nytimes.com/2018/07/08/business/china-surveillance-technology.html> accessed June 2019

[40] Roman Matzutt and others, 'MyneData: Towards a Trusted and Unser-Controlled Ecosystem for Sharing Personal Data', Gesellschaft für Informatik (2017)

[41] Guy Zyskind, Oz Nathan, Alex Pentland, 'Decentralizing Privacy: Using Blockchain to Protect Personal Data', 2015 IEEE Security and Privacy Workshops (2015); Lennart Bader and others, 'Smart Contract-Based Car Insurance Policies', *Proceedings of the 2018 IEEE Global Communications Conference (GLOBECOM)* (IEEE 2018)

[42] Sarah Meiklejohn and others, 'A Fistful of Bitcoins: Characterizing Payments Among Men with No Names', *IMC* (2013); Micha Ober, Stefan Katzenbeisser, Kay Hamacher, 'Structure and Anonymity of the Bitcoin Transaction Graph' (2013) 5 Future Internet 237; Fergal Reid, Martin Harrigan, 'An Analysis of Anonymity in the Bitcoin System' in Yaniv Altshuler and others (eds), Security and Privacy in Social Networks (2013); Jan Henrik Ziegeldorf and others, 'Secure and Anonymous Decentralized Bitcoin Mixing' (2018) 80 FGCS 448

[43] Roman Matzutt and others, 'A Quantitative Analysis of the Impact of Arbitrary Blockchain Content on Bitcoin', *Proceedings of the 22nd International Conference on Financial Cryptography and Data Security (FC) (*Springer 2018)

[44] Ibid.

## III. Information Sensitivity from a Legal Perspective

Since May 2018 the data protection law in the member states of the European Union is almost exclusively determined by the European General Data Protection Regulation (GDPR), leaving only a very limited scope of application to distinct national data protection legislation.[46] Depending on the content of the data, the European data protection law distinguishes between three different categories: personal, special categories of personal, and non-personal data, which are granted different levels of protection.

Art. 4 No. 1 GDPR defines *personal data* as any information relating to an identified or identifiable natural person. *Special categories of personal data* consist of personal data referring to particularly sensitive information concerning a natural person. Under Art. 9 sec. 1 GDPR these include data revealing a person's racial or ethnic origin, political opinions, religious or philosophical beliefs or union membership. Furthermore, genetic, biometric and health data as well as data concerning a person's sex life or sexual orientation are part of this category. Every piece of information not coming under the definition of personal data, however, has to be considered *non-personal data* under data protection law [cf. Art. 3 No. 1 Proposal for a Regulation of the European Parliament and of the Council on a framework for the free flow of non-personal data in the European Union, COM (2017) 495 final].

The level of protection of these different types of data varies widely. The protection of natural persons' personal data in relation to processing is a fundamental right under Artt. 7, 8 of the Charter of Fundamental Rights of the European Union as well as under national laws and constitutions of the member states, e.g., Art. 2 sec. 1 icw. Art. 1 sec. 1 GG. European and national Data protection law, therefore, provide various provisions to ensure this protection. Elementary to this concept is the principle of "ban with reservation of permit," which is laid down in Art. 6 sec. 1 cl. 1 GDPR.[47] It, only, allows data processing, if a law provides a legal basis for it.[48] The processing of personal data, furthermore, has to be conducted lawfully, transparently, and fairly (Art. 5 sec. 1 lit. a) GDPR). It is limited to a particular and determined purpose and the amount of data necessary for this (Art. 5 sec. 1 lit. b), c) GDPR). The data processed has to be accurate and may only be stored for the limited amount of time necessary for the purpose of the processing (Art. 5 sec. 1 lit. d), e) GDPR). At last, the processing has to ensure the integrity and confidentiality of the data (Artt. 5 sec. 1 lit. f), 32 GDPR). Further general duties of the data processors consist of ensuring data protection by the design of their products and its defaults (Art. 25 GDPR) as well as the conduction of a data protection impact assessment concerning their processing. The data protection law, moreover, grants data subjects several rights against the data processors. They hold the rights to transparent information about the data processing (Artt. 12 - 14 GDPR), the right to access information concerning the data stored about them (Art. 15 GDPR), the right to rectify wrong data stored (Art. 16 GDPR), and the right to erase data or restrict the further use of data (Artt. 17, 18 GDPR). All these rights and obligations are combined with efficient legal enforcement mechanisms, creating a high level of protection for personal data.

---

[45] Jan Henrik Ziegeldorf and others, 'Choose Wisely: A Comparison of Secure Two-Party Computation Frameworks', *2015 IEEE Security and Privacy Workshops* (2015)

[46] Heinrich Amadeus Wolff, Stefan Brink, Einleit. DSGVO, par 19 et seq. in Heinrich Amadeus Wolff, Stefan Brink (eds), *Beck`scher Onlinekommentar* Datenschutzrecht (25th edn, 2017)

[47] Marion Albers, Raoul-Darius Veit, Art. 9, par 4 in Heinrich Amadeus Wolff, Stefan Brink (eds), *Beck`scher Onlinekommentar Datenschutzrecht* (25th edn, 2018); Winfried Veil, 'Die Datenschutz-Grundverordnung: Des Kaisers neue Kleider' (2018) 37 Neue Zeitschrift für Verwaltungsrecht 686

[48] ibid.

The group of personal data summarized under the notion of "special categories" consists of types of personal data which can be described as sensitive personal attributes. They have in common that they concern very personal beliefs or states which bear a special risk of being a leverage point for discrimination[49] and are closely connected to the exercise of fundamental rights.[50] The processing of this data can result in a severe violation of person's privacy as well as significant risks to the fundamental rights and freedoms.[51] The legal protection for special categories of personal, therefore, has to be even stronger compared to common personal data. Following the principle of "ban with reservation to permit",[52] Art. 9 GDPR restricts the processing of special categories of personal data to less, more specific and more essential situations. Furthermore, only under very strict conditions personal data of the special categories may, at all, be used for decision making based solely on automated processing, including profiling (Art. 22 GDPR). If special categories of personal data are processed, this always leads to the necessity for the processor to keep records of his processing activities (Art. 30 GDPR). If handling this kind of data in a larger scale, the processor has to conduct

---

[49] Marion Albers, Raoul-Darius Veit, Art. 9, par 4 in Heinrich Amadeus Wolff, Stefan Brink (eds), Beck`scher Onlinekommentar Datenschutzrecht (25th edn, 2018); Jürgen Kühling, Benedikt Buchner, Art. 9, par 2 in Jürgen Kühling, Benedikt Buchner (eds), Datenschutz-Grundverordnung, Bundesdatenschutz: DS-GVO / BDSG (1st edn, 2018)

[50] Recital 51, cl. 1 GDPR; Martin Franzen, Art. 9, par. 1 DSGVO in Martin Franzen, Inken Gallner, Hartmut Oetker (eds), Kommentar zum Europäischen Arbeitsrecht (2nd edn, 2018); Marion Albers, Raoul-Darius Veit, Art. 6, par 2, Art. 9, par 4 in Heinrich Amadeus Wolff, Stefan Brink (eds), Beck`scher Onlinekommentar Datenschutzrecht (25th edn, 2018); Thilo Weichert, '"Sensitive Daten" Revisited' (2017) 41 Datenschutz und Datensicherheit 538

[51] Recital 51 cl. 1 GDPR; Martin Franzen, Art. 9, par. 1 DSGVO in Martin Franzen, Inken Gallner, Hartmut Oetker (eds), Kommentar zum Europäischen Arbeitsrecht (2nd edn, 2018)

[52] Marion Albers, Raoul-Darius Veit, 'Art. 6, par 2 in Heinrich Amadeus Wolff, Stefan Brink (eds), Beck`scher Onlinekommentar Datenschutzrecht (25th edn, 2018); Winfried Veil, 'Die Datenschutz-Grundverordnung: Des Kaisers neue Kleider' (2018) 37 Neue Zeitschrift für Verwaltungsrecht 686

data protection impact assessment (Art. 35 sec. 3 lit. b) GDPR) and is obliged to appoint a data protection officer (Art. 37 sec. 1 lit. c) GDPR).

As *non-personal data* does not fall under the scope of the fundamental rights which establish data protection, it is, hence, neither protected under the GDPR nor national data protection laws. Non-personal data, however, may be protected by other laws under different legal means, e.g., business secrets which are protected by the national civil law.[53]

The GDPR as well as the national data protection legislations differentiate between data protection and data security. While data protection aims at the defense of personal data against the dangers of their processing[54], data security embraces all measures to preserve data from misuse and interference of risks from outside of the process of processing.[55] A legal use of a personal data always requires adequate safety and security measures. An appropriate security level considers the technical state-of-the-art, the costs of the implementation of the security measures, the probability of occurrence of security risks, nature, scope, context and purposes of the processing as well as the risks for the rights and freedoms of the natural persons which might especially arise from the accidental or unlawful destruction, loss or unauthorized disclosure.[56] The processing of particular sensitive data may, hence, only lead to more data security, while in reverse the processing of common personal data does not mean to the absence of security measures.

## IV. Information Sensitivity from a User Perspective

From the user perspective, the perception of how sensitive an information is influences whether users feel their privacy to be preserved and thus, how concerned they are about said data provision or access. Information privacy can be defined as "the claim of individuals, groups or institutions to determine for themselves when, how, and to what extent information about them is communicated to others."[57] Research has shown that users are more concerned and less willing to provide information when it is perceived as highly sensitive.[58] For example, medical information is felt to be

---

[53] Dirk Müllmann, 'Auswirkungen Der Industrie 4.0 Auf Den Schutz von Betriebs- Und Geschäftsgeheimnissen' (2018) 64 Wettbewerb in Recht und Praxis 1177

[54] Heinrich Amadeus Wolff, Stefan Brink, §9 BDSG [2003] par.12 in Heinrich Amadeus Wolff, Stefan Brink (eds), Beck`scher Onlinekommentar Datenschutzrecht (25th edn, 2017)

[55] Heinrich Amadeus Wolff, Stefan Brink, §9 BDSG [2003] par.9 in Heinrich Amadeus Wolff, Stefan Brink (eds), Beck`scher Onlinekommentar Datenschutzrecht (25th edn, 2017); Hanns-Wilhelm Heibey, Chap. 4.5, par. 2 in Alexander Roßnagel (ed), Handbuch Datenschutzrecht (1st edn, 2003)

[56] cf. Art. 32 sec. 1, 2 GDPR

[57] Alan Westin, 'Privacy and Freedom.' (1967) 33 American Sociological Review 173 p 7

[58] David Mothersbaugh and others, 'Disclosure Antecedents in an Online Service Context: The Role of Sensitivity of Information' (2012) 15 Journal of Service Research 76; Shu Yang, Yuan Wang, Kan-Liang Wang, 'The Influence of Information Sensitivity Compensation on Privacy Concern and Behavioral Intention' (2009) 40 ACM SIGMIS Database 38; Gaurav Bansal, Fatemeh Zahedi, David Gefen, 'The Impact of Personal Dispositions on

sensitive,[59] especially information about psychological health.[60] Markos and colleagues[61] found that for US Americans the social security number is the most sensitive, whereas information about number of children, height, and race are perceived as the least sensitive. In their American study, medical information and location data -- that have been previously studied to be very sensitive[62] -- are evaluated to be not as sensitive as, e.g., financial data.

But what causes users to perceive certain information types as more sensitive than others?

The perception of sensitivity is related to the perceived risks when disclosing such information and, thus, related to the vulnerability and potential losses that are anticipated.[63] Users are concerned about unauthorized use, misuse (e.g., fraud, identity theft, hackers), and improper access.[64] But they also feel that the collection of information itself, targeted advertising, and profiling are violations of their privacy.[65] Thus, they seem not to differentiate between data privacy and data security. Empirical research also shows that more personally identifying information is perceived as more sensitive,[66] which goes in line with the GDPR covering personally identifying information. As we discussed in Section II. Information Sensitivity from a Technological Perspective, improving data analysis technologies enable ever-deeper insights about users. Most users may not be aware about what is

---

Information Sensitivity, Privacy Concern and Trust in Disclosing Health Information Online' (2010) 49 Decision Support Systems 138;

[59] Andrew J Rohm, George R Milne, 'Just What the Doctor Ordered The Role of Information Sensitivity and Trust in Reducing Medical Information Privacy Concern' (2004) 57 Journal of Business Research 1000; Gaurav Bansal, Fatemeh Zahedi, David Gefen, 'The Impact of Personal Dispositions on Information Sensitivity, Privacy Concern and Trust in Disclosing Health Information Online' (2010) 49 Decision Support Systems 138

[60] Andre Calero Valdez, Martina Ziefle, 'The Users' Perspective on the Privacy-Utility Trade-Offs in Health Recommender Systems' [2018] International Journal of Human-Computer Studies 108

[61] Ereni Markos, George R. Milne, James Peltier, 'Information Sensitivity and Willingness to Provide Continua: A Comparative Privacy Study of the United States and Brazil' (2017) 36 Journal of Public Policy & Marketing 79

[62] Andrew L. Rohm, George R Milne, 'Just What the Doctor Ordered The Role of Information Sensitivity and Trust in Reducing Medical Information Privacy Concern' (2004) 57 Journal of Business Research 1000; Gaurav Bansal, Fatemeh Zahedi, David Gefen, 'The Impact of Personal Dispositions on Information Sensitivity, Privacy Concern and Trust in Disclosing Health Information Online' (2010) 49 Decision Support Systems 138; Flavius Kehr and others, 'Blissfully Ignorant: The Effects of General Privacy Concerns, General Institutional Trust, and Affect in the Privacy Calculus' (2015) 25 Information Systems Journal 607

[63] David Mothersbaugh and others, 'Disclosure Antecedents in an Online Service Context: The Role of Sensitivity of Information' (2012) 15 Journal of Service Research 76; May Lwin, Jochen Wirtz, Jerome D Williams, 'Consumer Online Privacy Concerns and Responses: A Power–Responsibility Equilibrium Perspective' (2007) 35 Journal of the Academy of Marketing Science 572

[64] European Commission, 'Data Protection Eurobarometer' (2015); H. Jeff Smith, Sandra J. Milberg, Sandra J. Burke, 'Information Privacy: Measuring Individuals' Concerns about Organizational Practices' (1996) 20 Management Information Systems Quarterly 167

[65] Eva-Maria Schomakers, Chantal Lidynia, Martina Ziefle, 'Hidden within a Group of People Mental Models of Privacy Protection', 3rd International Conference on Internet of Things, Big Data and Security (SCITEPRESS - Science and and Technology Publications 2018); H. Jeff Smith, Sandra J Milberg, Sandra J. Burke, 'Information Privacy: Measuring Individuals' Concerns about Organizational Practices' (1996) 20 Management Information Systems Quarterly 167

[66] Miguel Malheiros, Sören Preibusch, M. Angela Sasse, '"Fairly Truthful": The Impact of Perceived Effort, Fairness, Relevance, and Sensitivity on Personal Data Disclosure' (2013) 7904 LNCS 250

legally and technically possible and how presumably non-identifiable or insensitive data can be linked and used.[67]

Besides limited knowledge about IT and law there is another aspect, that complicates the sensitivity evaluation for the users: Privacy perceptions, desires, and perceived privacy risks depend on context and audience.[68] In a medical consultation, we may disclose medical information to a stranger (the physician) without worries, which we would not share with good friends. We learn from early childhood on, how to manage our privacy in an offline world, e.g., by closing doors and holding conversations in inaudible distance from unwanted listeners. But online, data is persistently available over space and time, confusing the context in which we disclose information and those, in which it can be accessed and by whom.[69] With the collapsing contexts of data disclosure in the online world due to the persistence and replicability of data,[70] users do not only have to include the present audience and context to evaluate the risk of disclosure and sensitivity of information but also potential access of information in the future by different entities and in different contexts. And the technological possibilities to combine data across services also need to be considered. At this point the question arises, whether the implicit risk evaluation users apply to determine their perception of sensitivity of information coincides with what is legal and what is technically possible.

## 1. The Empirical Approach

To provide a European user perspective on sensitivity of information, we conducted an empirical study in which n = 601 German internet users evaluated 40 data types regarding their felt sensitivity in an online questionnaire. The perceived sensitivity is assessed without contextual frame, as in the online world, contexts, audiences, and time frames melt together. Thus, such context-free perceptions of sensitivity mirror assessments users need to make in the digital world.

The online questionnaire started with a short introduction to the topic and demographic questions (age, gender, education level). In the main part, the participants evaluated 40 information types on a 6-point scale from "not sensitive at all`` (1) to "very sensitive`` (6). 32 information types were taken

---

[67] European Commission, Flash Eurobarometer 443: Report e-Privacy (2016)

[68] Helen Nissenbaum, 'Privacy In Context: Technology Policy And The Integrity Of Social Life' 304; Irwin Altman, 'Privacy - A Conceptual Analysis' (1976) 8 Environment and Behavior 7

[69] Leysia Palen, Paul Dourish, 'Unpacking "Privacy" for a Networked World' [2003] Proceedings of the conference on Human factors in computing systems - CHI '03 129; Monika Taddicken, 'The "Privacy Paradox" in the Social Web: The Impact of Privacy Concerns, Individual Characteristics, and the Perceived Social Relevance on Different Forms of Self-Disclosure1' (2014) 19 Journal of Computer-Mediated Communication 248

[70] Monika Taddicken, 'The "Privacy Paradox" in the Social Web: The Impact of Privacy Concerns, Individual Characteristics, and the Perceived Social Relevance on Different Forms of Self-Disclosure1' (2014) 19 Journal of Computer-Mediated Communication 248

from the study of Markos et al.[71] for comparability. Eight ones were chosen as they have been controversially discussed by participants of a preceding focus group study. To prevent sequence effects, the order of data types was randomized for each participant. The data types are listed in Figure 1.

## 2. The Sample

601 participants between 15 and 69 years of age completed the online questionnaire (M = 38.8, SD = 20.23). 59.1% were women. The questionnaire was distributed online via an independent market research company. The education level is also quite heterogeneously distributed (cf. Table 1) showing a good cross-section of German internet users.

Table 1. Demographic characteristics of the sample, n = 601.

|  |  | distribution in sample (n = 601) |
|---|---|---|
| age [years] | mean (SD) | 38.8 (20.23) |
|  | min – max | 15 - 69 |
| gender | women | 59.1% |
|  | men | 40.9% |
| education | no certificate | 7.2% |
|  | certificate of secondary education | 26.3% |
|  | apprenticeship | 19.3% |
|  | qualification for university entrance | 26.1% |
|  | university degree | 21.1% |

## 3. Users Evaluation of Data Sensitivity

The perceived sensitivity for all 40 data types is depicted in Figure 1. On average, passwords are felt to be the most sensitive information type out of the presented ones ($M$ = 5.57, $SD$ = 0.94), followed by financial account numbers ($M$ = 5.55, $SD$ = 0.975). Both are evaluated as "very sensitive" on average with the mean higher than 5.5. Data types that are perceived to be *"sensitive"* (4.5 < $M$ < 5.5) range from personal identifiers like passport number ($M$ = 4.99, $SD$ = 1.33) and fingerprint ($M$ = 4.94, $SD$ = 1.45) to GPS location ($M$ = 4.63, $SD$ = 1.34), home address ($M$ = 4.61, $SD$ = 1.47), and medical history ($M$ = 4.6, $SD$ = 1.37). Users perceive as "rather sensitive" (3.5 < $M$ < 4.5) for example the following: online activities like browsing history ($M$ = 4.33, $SD$ = 1.35), law enforcement files ($M$ = 4.39, $SD$ = 1.71), medication ($M$ = 4.04, $SD$ = 1.54), sexual preferences ($M$ = 3.84, $SD$ = 1.77), and license plate number ($M$ = 3.64, $SD$ = 1.58). "*Rather not sensitive*" are, for example, political affiliation ($M$ = 3.38, $SD$ = 1.45), weight ($M$ = 3.26, $SD$ = 1.53), zip code ($M$ = 3.13, $SD$ = 1.45), occupation ($M$ = 2.85, $SD$ = 1.37), and height (*M* = 2.51, *SD* = 1.29). The only two information types from this list that are perceived as *"not sensitive"* (1.5 < M < 2.5) are hair color (*M* = 2.19, *SD* = 1.33) and name of pet (*M* = 2.09, *SD* = 1.31). None is on average felt to be *"not sensitive at all"* (*M* < 1.5).

---

[71] Ereni Markos, George R. Milne, James W Peltier, 'Information Sensitivity and Willingness to Provide Continua: A Comparative Privacy Study of the United States and Brazil' (2017) 36 Journal of Public Policy & Marketing 79

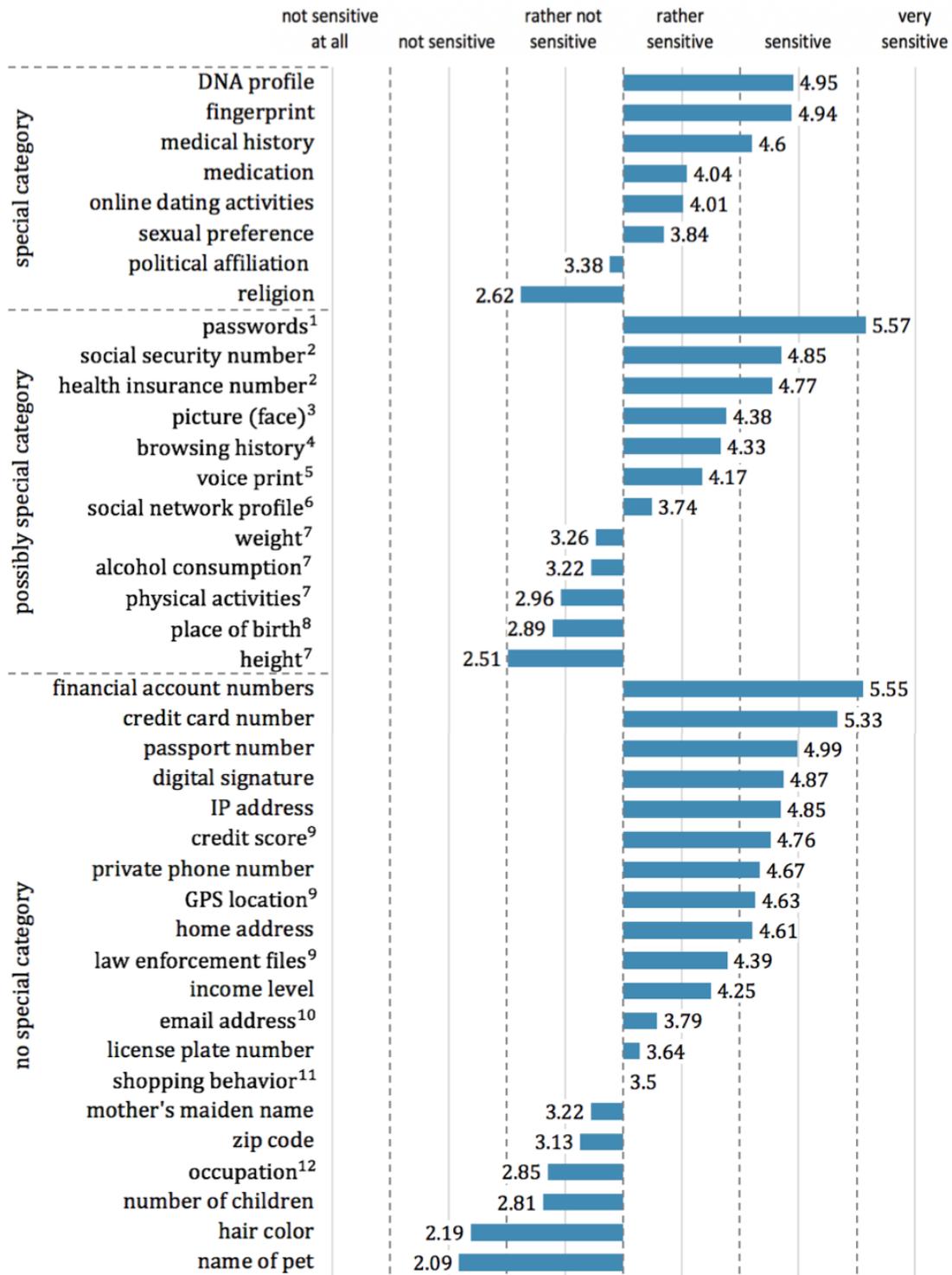

Figure 1. Users' evaluation of the sensitivity of 40 data types (n = 601) categorized into the legal classification.

## V. Comparison between Legal, Technical, and User Perspective on Information Sensitivity & Discussion

After presenting the technological, legal, and user perspective on information sensitivity separately, we will in this chapter compare the three perspectives and discuss the differences and similarities. Figure 1 depicts the legal and user perspective on sensitivity. It shows that users' perception of sensitivity is for some data types in line with the legal categorization, but also deviates strongly for other data types. The data types that are perceived as most sensitive by the users (passwords, financial account numbers) are legally classified as possibly special category and no special category. This assessment by the users indicates that they do not differentiate between data privacy and data security. Rather, the sensitivity evaluation is based on a risk assessment, and users are concerned about unauthorized access and illicit data misuse as well as about data collection, targeted advertising, and profiling (cf. Section IV. Information Sensitivity from a User Perspective). The legal use of personal data, however, under Art. 32 GDPR always requires adequate data security measures proportional to the risks of data processing (cf. Section III. Information Sensitivity from a Legal Perspective). Hence, service operators implement established technical protection measures, e.g., by especially protecting user passwords that need to be stored on their facilities. However, as we discussed in Section II. Information Sensitivity from a Technological Perspective, even despite huge efforts to technically protect such user data, we experience frequent data breaches, as is documented by the service portal ``Have I Been Pwned?''.[72]

Data protection, however, starts earlier and already tries to minimize the occasions and purposes in which personal data is collected and processed. While the users' estimation of the data sensitivity might anticipate the uncontrolled release and accept the necessity of the processing in other contexts, the data protection law works with wider categories and has internalized the dependence of the context. Hence, from a legal point of view the additional protection of special categories of personal data aims at categories whose general acquaintance, irrespective of its legality, might bear severe risks and consequences. Therefore, the legislator even restricted the contexts of legal uses compared to regular personal data.

Political affiliation and religion are classified as special category and thus, particularly deserving protection by the GDPR, but are assessed as `rather not sensitive' on average by the participants of the survey. The German view on data protection is, among other factors, highly influenced by the country's historical experience of two dictatorships cementing their power through surveillance and control of the citizens and the potential as well as the risks of modern electronic data processing.[73] This influenced the development of the European data protection law.[74] As illustrated in Section III. Information Sensitivity from a Legal Perspective, the legal point of view on special categories of personal data, as the most sensitive pieces of information in data protection law, mainly concern issues that can be used as leverage points for discrimination, such as religious believes, political opinions, or sexual orientation, and are closely connected to the exercise of fundamental rights, e.g., union memberships. Therefore, they need the particular protection of the democratic society and its laws. This aspect does not have the same importance for common users and the deviation between law and user evaluation can be explained by the methodological approach to report the *mean* user evaluation. For many users who have a mainstream political attitude or a religion that is not discriminated, these information types may not seem sensitive. The minority of those users who may be discriminated for these characteristics do not have much weight within the average evaluation, but still need to be protected from discrimination. Similarly, sexual preferences are probably more than "rather sensitive" for some but not all internet users. As mentioned above short-term financial losses and other acute consequences of released data are more relevant in the users' world of experience, while the legislator has to consider long-term implications for the individual and the democratic society as a whole.

---

[72] Troy Hunt, 'Have I Been Pnwed?' (2013) <https://haveibeenpwned.com> accessed June 2019

In Section 2. Increasing Data Sensitivity, we have seen that the voluntary sharing of personal information in social media can make users vulnerable and create possibilities for harm, as showcased by the potential information leakage due to meta data or advanced analyses such as performed by Cambridge Analytica. Users see these risks to some extend and state that they are concerned, but still disclose this information.[75] One explanation for this privacy paradoxical user behavior is given by the theory of the privacy calculus,[76] which assumes that users weigh perceived benefits and perceived privacy risks against each other. Thus, they disclose information when the benefits outweigh the risks. Correspondingly, the evaluation of risks is only one side of the coin. Self-disclosure on social network

---

[73] Hans Peter Bull, Informationelle Selbstbestimmung - Vision oder Illusion? (2nd edn, Mohr Siebeck 2011); Johannes Masing, 'Herausforderungen des Datenschutzes' (2012) 65 Neue Juristische Wochenschrift 2305

[74] Viviane Reding, 'Sieben Grundbausteine Der Europäischen Datenschutzreform' (2012) 2 Zeitschrift für Datenschutz 195

[75] Nina Gerber, Paul Gerber, Melanie Volkamer, 'Explaining the Privacy Paradox: A Systematic Review of Literature Investigating Privacy Attitude and Behavior' (2018) 77 Computers and Security 226;

[76] Nina Gerber, Paul Gerber, Melanie Volkamer, 'Explaining the Privacy Paradox: A Systematic Review of Literature Investigating Privacy Attitude and Behavior' (2018) 77 Computers and Security 226; Tamara Dinev, Paul Hart, 'An Extended Privacy Calculus Model for E-Commerce Transactions' (2006) 17 Information Systems Research 61

sites bring many benefits to the individual including self-representation, relationship development, and social control.[77] These may outweigh the perceived concerns.

Additionally, research has shown that users do not make a purely rational decision. Rather, decision making is affected by cognitive biases and heuristics.[78] For example, optimism bias lead individuals to the perception that they themselves are less vulnerable than others[79] and affect heuristics influences the risk assessment in a way that users tend to underestimate risks and correspondingly disclose more, when it is associated with things they like.[80] These psychological means will always influence users' decision making to some extent.

Here, the aim of data legislation and privacy preserving technologies should be to guarantee users an online environment in which they can freely decide what to share, in line with the principle of informational self-determination. This also includes data protection via technical and legal means, so that users are protected to the largest extent possible. This is particularly important as the online context is marked by persistence of information, making it harder for users to delete once published data.

Another deviation between users' and legal evaluation is the categorization of location information. GPS data can either comprise distinct locations or even whole trajectories as discussed in Section 2. Increasing Data Sensitivity and is felt to be sensitive by users. But location is not part of the group of special categories of data under Art. 9 GDPR. Nevertheless, there is European legislation providing special rules for its processing. Directive 2002/58/EC, which was enacted to complement the former European Data Protection Directive 95/46/EC, defines it as any data processed in an electronic communications network, indicating the geographic position of the terminal equipment of a user of a publicly available electronic communication service. (Art. 2 lit. c) dir. 2002/58/EC) Art. 9 sec. 1 dir. 2002/58/EC allows the processing of this data only after its anonymization or with the consent of the users to the extent and for the duration necessary to provide an additional service. The scope of this provision, however, is limited to the regulation of data processing in the context of providing publicly available electronic communications networks (cf. Art. 3 sec. 1 dir. 2002/58/EC),

---

[77] Haein Lee, Hyejin Park, Jinwoo Kim, 'Why Do People Share Their Context Information on Social Network Services? A Qualitative Study and an Experimental Study on Users' Behavior of Balancing Perceived Benefit and Risk' (2013) 71 International Journal of Human-Computer Studies 862; Saadi Lahlou, 'Identity, Social Status, Privacy and Face-Keeping in Digital Society' (2008) 47 Social Science Information 299

[78] Alessandro Acquisti, Laura Brandimarte, George Loewenstein, 'Privacy and Human Behaviour in the Age of Information' (2015) 347 Science 509}

[79] Hichang Cho, Jae-Shin Lee, Siyoung Chung, 'Optimistic Bias about Online Privacy Risks: Testing the Moderating Effects of Perceived Controllability and Prior Experience' (2010) 26 Computers in Human Behavior 987

[80] Flavius Kehr and others, 'Blissfully Ignorant: The Effects of General Privacy Concerns, General Institutional Trust, and Affect in the Privacy Calculus' (2015) 25 Information Systems Journal; Wakefield R, 'The Influence of User Affect in Online Information Disclosure' (2013) 22 Journal of Strategic Information Systems 157

e.g., by phone companies or closed user groups.[81] For every other purpose and processing the rules of the GDPR apply subsidiarily, treating location data connected to person as regular personal data. Hence, only in a very limited number of use cases relevant today a further protection of location data is granted by law.

Considering the possibility of creating movement profiles of users through the analysis of location data and the threats to a person's freedom and rights such profiles bear, the protection granted by law seems insufficient. One reason for this situation is the ongoing reform process of the European data protection law. The directive 2002/58/EC relevant at hand will be renewed and transferred into a regulation in the near future (cf. Proposal for as Regulation concerning the respect for private life and the protection of personal data in electronic communications and repealing Directive 2002/58/EG, COM 2017/010final-2017/03(COD)). Hence, the current legal state does not, yet, meet today's technical challenges. Despite of the oncoming reform, the classification of location data as common personal data within the GDPR should be reconsidered and a higher level of protection for location data should be created.

The contrast between the European law and examples like the "Chinese Social Credit System"[82] shows that, in other cases, due to sufficient legal regulations in Europe, users are protected from harmful aggregation of data, although this would be technically possible. The collection of license plate numbers to create a governmental "obedience score" is not conceivable, as the European fundamental rights exclude, e.g., the comprehensive assessment of a person's behavior and actions. Therefore, European users do not need to be concerned about possible consequences which could result from the collection of such data. The low sensitivity evaluation of license plate number by the German sample shows that users indeed do not see many risks connected to that information. This indicates that either they rely on the law to protect them or that they are not aware of the technological possibilities to use this type of information (against them).

The latter assumption -- users not being aware of potential risks -- is important to consider regarding users' perception of privacy risks in general. Looking at data disclosure decisions through a privacy calculus lens, users need to evaluate the risks of data disclosure. To do that sufficiently, they need to be aware of these risks. However, they are not always privy to the legal protection and technical means.[83]

---

[81] Peter Büttgen, §91 TKG par.16 in Klaus-Dieter Scheurle, Thomas Mayen (eds), Telekommunikationsgesetz (3th edn, 2018); Jens-Daniel Braun, §91 TKG par.9 in Martin Geppert, Raimund Schütz (eds), Beck`scher TKG Kommentar (4th edn, 2013)

[82] Mirijam Meissner, 'China's Social Credit System: A Big-Data Enabled Approach to Market Regulation with Broad Implications for Doing Business in China' (2017) 24 Mercator Institute for China Studies

[83] Sabine Trepte and others, 'A Cross-Cultural Perspective on the Privacy Calculus' (2017) 3 Social Media + Society; European Comission, Flash Eurobarometer 443: Report e-Privacy (2016)

The authors argue that the main objective should not be informational heteronomy that is imposed by law over the users but informational self-determination. But this requires that users are aware and able to evaluate the risks of data disclosures. As we see, users are in fact not knowledgeable about the legal protection and the technical possibilities to use and combine data, this awareness is oftentimes missing. To empower users here, they need to be well informed. But educational measures at school are limited, especially because of the ever-evolving technical means. Thus, education and information must be available from a trusted source for all citizens, e.g., from a governmental website. Also, qualified media coverage and easy to understand consent forms are required. Finally, technological means to prevent unauthorized access to user data or the derivation of additional information from such data need to be further improved, especially given the increasing complexity of how such data is being processed.

In summary, the comparison of the different perspectives on data sensitivity shows that users', technical, and legal views deviate to some degree. For example, the law grants protection to data categories as "special" categories that not all users perceive as especially sensitive. This can be seen as unproblematic as users are still able to freely disclose data, thereby giving their explicit consent to process this data. Other data categories, e.g., GPS data, are not given special privacy protection by the GDPR, but they are perceived as sensitive by the users and are, from a technological perspective, very revealing about the individual user. Here, the law nevertheless requires adequate data security measures for data processing, thus still providing protection. Rather, it is a problem that users, by allowing the processing of their data on the basis of consent without being fully aware of possible consequences, often thwart the safeguards of data protection laws which generally tries to limit the amount of processed data and the admissible purposes of its processing. For the premise of informational self-determination, it is of utmost importance to raise users' awareness about possible risks of data disclosure, on the one hand, and the legal protection they are entitled to, on the other hand. As long as the users decide to give their free, specific, informed, and unambiguous consent to the processing of their data as demanded by Artt. 7 sec. 1, 4 No. 11 GDPR the processing is in accordance with the law. It is, therefore, a manifestation of the user's informational self determination which would, otherwise, turn into informational heteronomy. Finally, technological means must seek to unburden the users, i.e., provide the best data protection possible while not overly restricting the users' freedom of educated self-expression. From a legal point of view, the different perception of data sensitivity by users, computer science, and law seems unproblematic. It would be an issue if the legal framework could not provide the data protection and security users are legitimately expecting or if the law would ignore threats to privacy uncovered by technical sciences. In the current state, admittedly, the perception of the importance between the three perspectives differs. However, the categories of data seen as sensitive by users and science are adequately protected by law, although it grants other categories, which are not rated as particularly sensitive by users or computer science, more protection for the aforementioned reasons.